\documentclass[12pt]{article}
\usepackage{graphicx}

\newcommand\pubdate{\today}

\def\roma{INFN Roma\\
Piazzale Aldo Moro, 2 I-00185 Roma Italy}

\def\Title#1{\begin{center} {\Large #1 } \end{center}}
\def\Author#1{\begin{center}{ \sc #1} \end{center}}
\def\Address#1{\begin{center}{ \it #1} \end{center}}

\newcommand\pubblock{\rightline{\begin{tabular}{l}  \\
         \pubdate  \end{tabular}}}
\newenvironment{Abstract}{\begin{quotation}  }{\end{quotation}}
\newenvironment{Presented}{\begin{quotation} \begin{center} 
             PRESENTED AT\end{center}\bigskip 
      \begin{center}\begin{large}}{\end{large}\end{center} \end{quotation}}
\def\Acknowledgements{\bigskip  \bigskip \begin{center} \begin{large}
             \bf ACKNOWLEDGEMENTS \end{large}\end{center}}
             
\def\muegamma{ $\mu \rightarrow e \gamma$ \,}
\newcommand*{\BR}     { {\cal B} }

\newcommand*{\megsign}        {\mathrm{\mu}^+ \to \mathrm{e}^+ \mathrm{\gamma}}

\newcommand*{\egamma}         {E_{\mathrm{\gamma}}}
\newcommand*{\epositron}      {E_\mathrm{e}}
\newcommand*{\tegamma}        {t_{\mathrm{e \gamma}}}

\newcommand*{\thetaegamma}    {\theta_{\mathrm{e \gamma}}}
\newcommand*{\phiegamma}      {\phi_{\mathrm{e \gamma}}}
\newcommand*{\thetae}    {\theta_\mathrm{e}}
\newcommand*{\phie}      {\phi_\mathrm{e}}

\begin{document}
\begin{titlepage}
\pubblock

\vfill
\Title{Searching for the \muegamma decay with MEG and MEG-II }
\vfill
\Author{ Gianluca Cavoto }
\Address{\roma}
\vfill
\begin{Abstract}
 The MEG collaboration searches  for the  \muegamma decay at the muon beam line  $\pi$E5  of the  Paul Scherrer Institut in Switzerland.
  The world best upper limit has been set to be $\BR$(\muegamma)  $<$  5.7 10$^{-13}$ at 90 \% C.L. analyzing a data set of 3.6 10$^{14}$  stopped muons on target.
  
   An upgrade program  of the detector (MEG-II) should lead to  a sensitivity at a level of  few 10$^{-14}$ within the next five years.

\end{Abstract}
\vfill
\begin{Presented}
FPCP 2014\\
Marseille, France,  May 25$^{th}$-30$^{th}$ 2014
\end{Presented}
\vfill
\end{titlepage}
\def\thefootnote{\fnsymbol{footnote}}
\setcounter{footnote}{0}

\section{Introduction}

 Lepton Flavour Violation in the decay of charged leptons  (cLFV) is of the utmost importance to find evidence of New Physics  (NP) beyond the Standard Model (SM) of fundamental interactions. The \muegamma decay in SM is predicted to have a tiny branching fraction ($\BR$) of the order of 10$^{-55}$, that is basically explained by the neutrino flavour oscillation and by the very small ratio of the neutrino  over the $W$ mass  \cite{petcov}. 
 
 At the same time  in almost every extension of the SM predicting  NP effects \cite{barbieri, hisano, LFV-EPC, isidori},  a $\BR$(\muegamma) at the level of 10$^{-12}$  is well possible,  calling for an  experimental effort to search for this decay.
  The    \muegamma  process  has a clear two body decay topology and advances in this search have been  obtained in the last  60 years with more and more precise detectors and  with  higher and higher  intensity muon beams.

\section{The MEG experiment}

 The MEG experiment has been  operated at the $\pi$E5  beam line of the Paul Scherrer Institut (PSI) at Villigen (CH) since 2007 until 2013.  A  continuous positive muon beam is available with a maximal rate of about 10$^8$ muon per second. MEG took data with an optimal muon stopping rate on a thin target of 3 10$^7$ muon/s.
 
  The MEG detector includes a  spectrometer  made of  a special gradient magnetic field  and low mass drift chamber planes to track the positrons emerging from the target. The magnetic field sweeps out of the tracking volume the lower momentum positrons and direct the higher momentum positrons on two fast scintillator arrays  readout  by PMT to measure the positron emission time.  Photons  produce showers in a  900 liter liquid Xe  calorimeter  where scintillating light is produced and detected by PMTs with a photon detection efficiency of about 60\% \cite{Adam:2013vqa}.

\subsection{MEG datasets and analysis}
  MEG collected data  until 2013 for a total number of  about 7 10$^{14}$ muon stopped on target. The results presented here are based on half of this data-set (collected in the 2009-2011 period) \cite{lastprl}. 

The signature of the signal event is given by a back-to-back, monoenergetic, time coincident 
photon-positron pair from the two body $\megsign$ decay.
In each event, positron and photon candidates are described by five 
observables: the photon and positron energies ($\egamma$, $\epositron$), 
their relative directions ($\thetaegamma$, $\phiegamma$)\,\cite{angledef}
and emission time ($\tegamma$). 
The  analysis is based on a maximum likelihood technique
applied in the analysis region defined by
$48<\egamma<58\,$MeV, $50<\epositron<56\,$MeV,
$\left|\tegamma\right|<0.7\,$ns,
$\left|\thetaegamma\right|<50\,$mrad and
$\left|\phiegamma\right|<50\,$mrad, which is described in 
detail in \cite{MEG2011}.

The background has two components, one coming 
from the Radiative Muon Decay $\mathrm{\mu}^{+}\rightarrow{\rm e}^{+} 
\mathrm{\nu} \bar{\mathrm{\nu}} \mathrm{\gamma}$ (RMD) and another from the accidental superposition 
of energetic positrons from the standard muon Michel decay with photons 
from RMD, positron-electron annihilation-in-flight or bremsstrahlung. 
At the MEG data taking rate, 93\% of events with $E_{\mathrm{\gamma}}>48~{\rm MeV}$ are from the ACCidental background (ACC).

\subsection{MEG results}
A blind analysis procedure is applied  by masking a region of 
$48<\egamma<58$\,MeV and $\left| \tegamma
\right|< 1\,{\rm ns}$
until the Probability Density Functions (PDFs) for the likelihood function are finalized. 
The  background studies and the extraction of the PDFs are carried out in data sidebands close to the signal region. Signal PDF are obtained from various sources:  monochromatic 55\,MeV photons from $\mathrm{\pi}^{0}$ decays  for $\egamma$, a fit to the  Michel positron $\epositron$ spectrum edge for $\epositron$,  $\tegamma$ in RMD events with  $\egamma$ $<$48 MeV for $\tegamma$ and special samples of tracks to validate the $\thetaegamma$ and  $\phiegamma$ PDFs.
 The performances obtained with the MEG detector are summarized by  the  observables resolution reported  in the first column of  Table \ref{tab:scenario}.

The maximum likelihood fit is performed in order  
to estimate the number of signal, RMD and ACC events in the analysis region.
The definition of the likelihood function is described in detail in \cite{MEG2011}.

\begin{figure}[h]
\centering
\includegraphics[height=8cm]{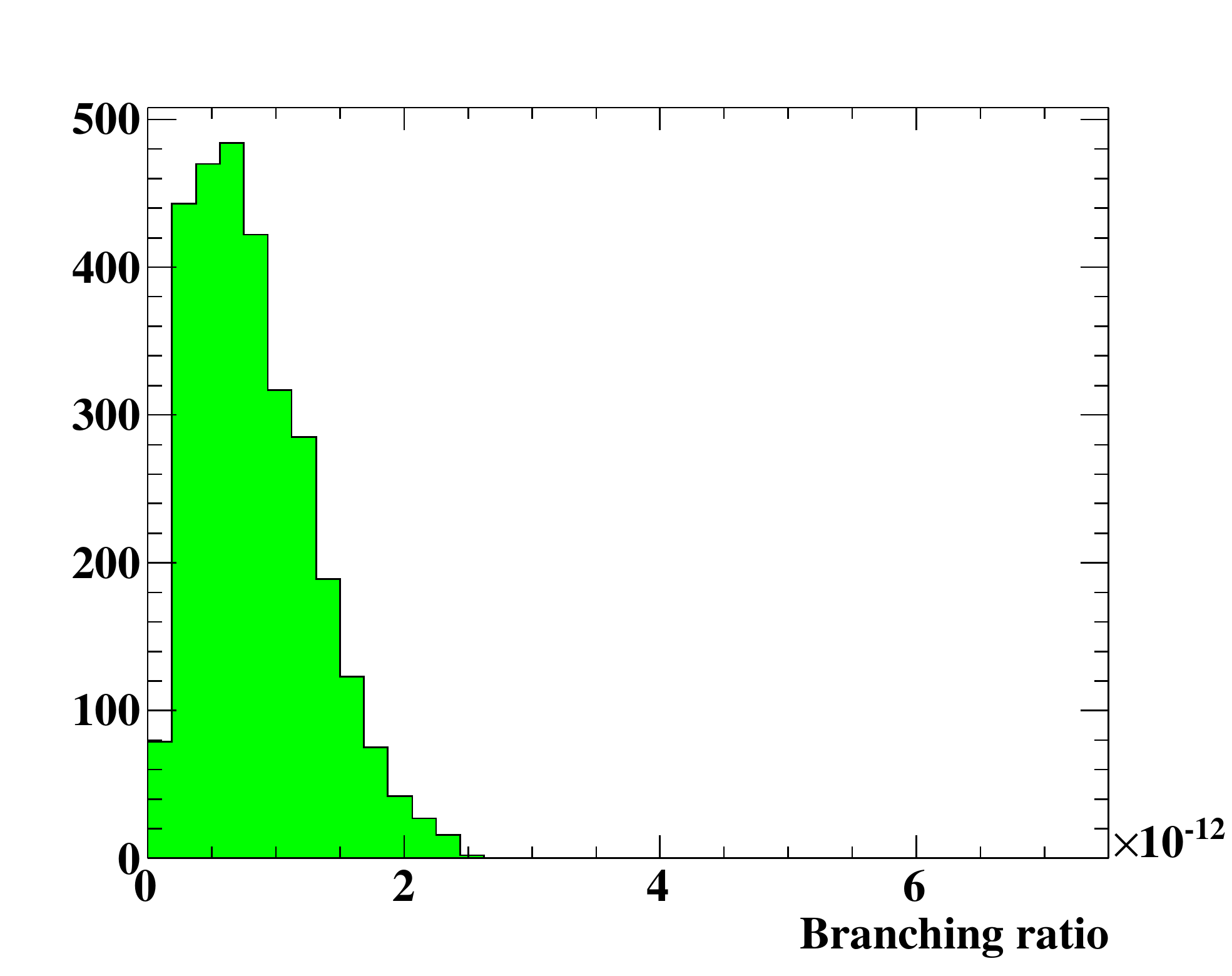}
\caption{Distribution of the  $\BR$  upper limits at $90\%$ C.L. for a  pseudo-experiments ensemble. }
\label{fig:sens}
\end{figure}

The distribution of  the $\BR$  upper limits at $90\%$ C.L. is 
obtained over an ensemble of pseudo-experiments, randomly 
generated according to the PDFs based on a 
null signal hypothesis, with the rates of ACC and RMD 
evaluated from the sidebands.  
The sensitivity  is estimated as the median of such distribution 
 (Fig.\ref{fig:sens}) to be 7.7 10$^{-13}$.

 No signal events are found in the signal region while  obtaining  $N_{\mathrm{RMD}}$ = 163 $\pm$ 32 and $N_{\mathrm{ACC}}$ = 2411 $\pm$ 57,

\begin{figure}[htb]
\centering
\includegraphics[height=8cm]{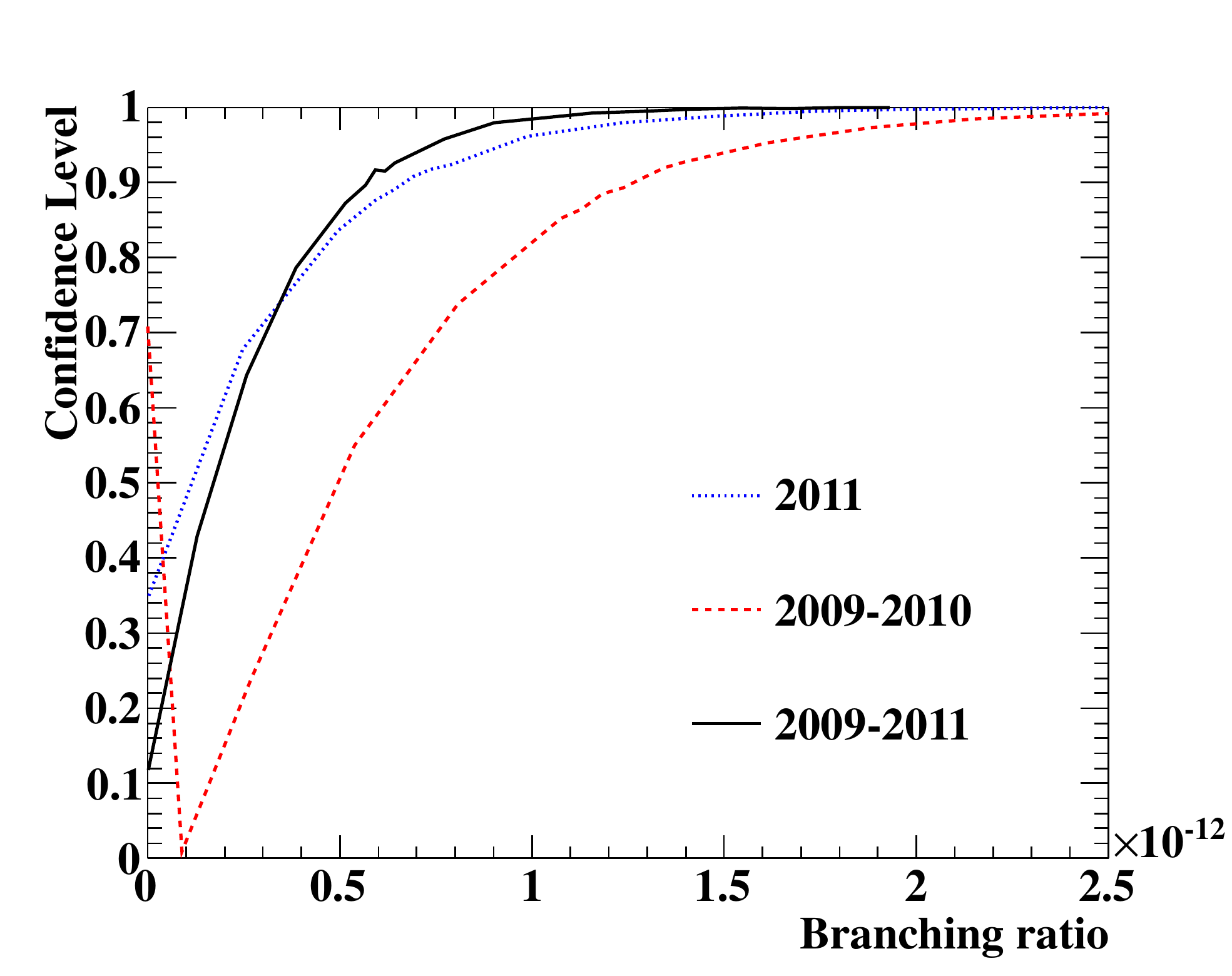}
\caption{Confidence level curve for the $\BR$ for the 2009-2011 data-set (and separately for the 2009-2010 and 2011 subsets).}
\label{fig:CL}
\end{figure}

The confidence interval for the number of signal events is calculated by a 
frequentist method with a profile likelihood-ratio 
ordering\,\cite{MEG2011, PDG, Feldman-Cousins}, 
where the numbers of RMD and ACC events are treated as nuisance parameters.

To translate the estimated number of signal events into a signal branching ratio two  methods 
are used, either counting the number of Michel positrons selected with a dedicated trigger or the number of RMD events observed in the muon data  leading to a 4\% uncertainty in the branching ratio estimate (Fig.\ref{fig:CL}).

The upper 
limit on $\BR$($\megsign$) is  
${\cal{B}} < 5.7\times 10^{-13}$ at $90\%$ C.L. which improves the previous best upper limit \cite{MEG2011} by a factor of four.

\subsection{MEG outlook} 

\begin{figure}[htb]
\centering
\includegraphics[height=8cm]{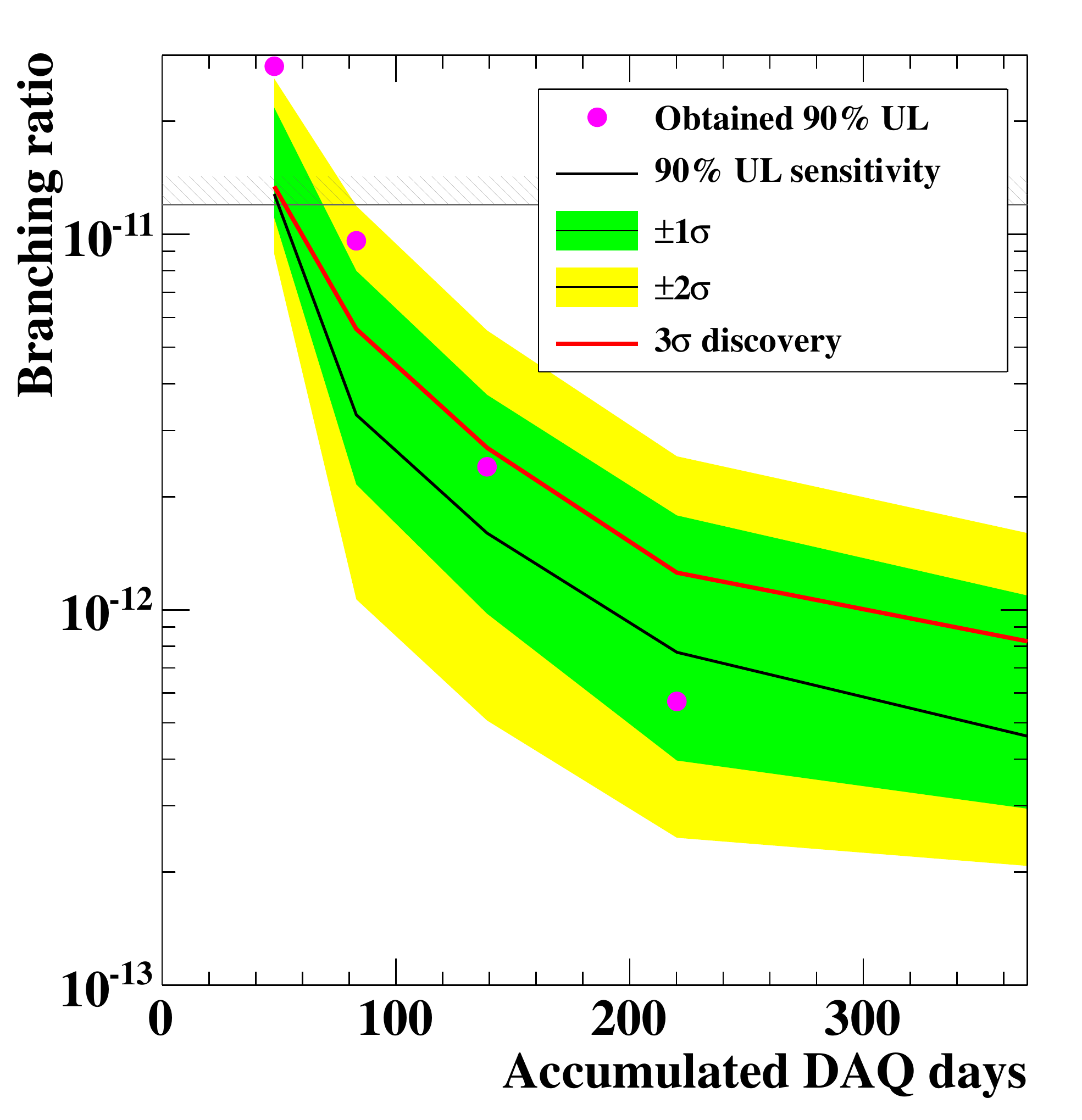}
\caption{Sensitivity of the current MEG detector as a function of the accumulated data.}
\label{fig:CLproj}
\end{figure}

   The final MEG results will use a data set that will be twice larger than the currently analyzed sample, with a corresponding predicted sensitivity of 5.0 $\times$ $10^{-13}$. Given the detector performances the sensitivity does not significantly improves adding more data (Fig.\ref{fig:CLproj}). 
   
   The dominant accidental background  (proportional to the square of the muon stopping  rate and to the various observables resolutions) can be reduced only with a better detector. This then allows to use a higher muon stopping rate. Moreover, several models of NP still leave room for $\BR$ in the $10^{-14}$-$10^{-13}$ range \cite{Antusch:2006vw}. 
     Given the availability of  a higher muon flux at PSI,  an upgrade program for the MEG detector in the next coming years is very desirable.

\section{MEG upgrade,  MEG-II}
  
   The MEG collaboration has recently proposed an upgrade program  (MEG-II) for its detector \cite{Baldini:2013ke} that has been approved by PSI and  by its funding agencies.  
 
\subsection{MEG-II proposal and sensitivity}

\begin{figure}[htb]
\centering
\includegraphics[height=8cm]{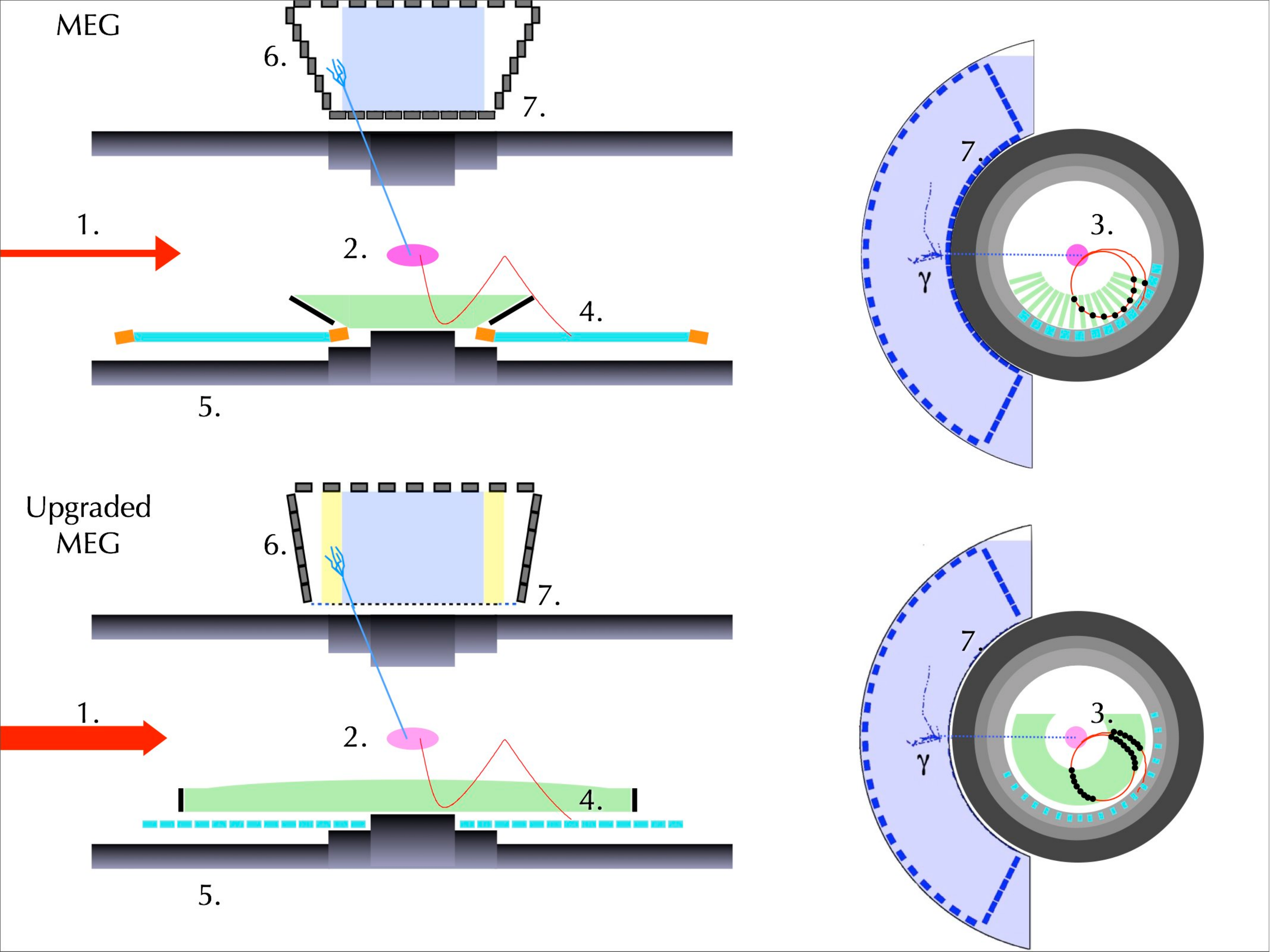}
\caption{An overview of the present MEG experiment versus the proposed upgrade. The 
numbers refer to the items listed in the text.}
\label{fig:upgrade}
\end{figure}

The MEG-II  relies on the following improvements compared with the present MEG experiment, shown 
schematically in Figure~\ref{fig:upgrade} and here enumerated:
\begin{enumerate}
\item Increasing the number of stopping muons on target;
\item Reducing the target thickness to minimize the material traversed by photons and positrons on their trajectories
towards the detector;
\item Replacing the positron tracker,  reducing its radiation length and improving its granularity and
resolutions;
\item Improving the positron tracking and timing integration, by measuring the $e^+$ trajectory  to the positron timing counter interface;
\item Improving the positron  timing counter granularity for better timing and reconstruction;
\item Extending the calorimenter  acceptance;
\item Improving the photon energy, position and timing resolution for shallow events;
\item Integrating splitter, trigger and DAQ while maintaining a high bandwidth.
\end{enumerate}

  This would lead to better resolutions as outlined in Table \ref{tab:scenario}. In particular a higher efficiency positron spectrometer with $\epositron$,  $\thetae$ and $\phie$  smaller resolutions would considerably improve the performance of the detector. With sizeable improvements in the other sub-detectors and with a muon stopping rate 7 $10^7$ muons per second MEG-II will have a predicted sensitivity of 5 $10^{-14}$  after 3 years of nominal data-taking.

\begin{table}[t]
\begin{center}
\caption{ \label{tab:scenario}Resolution (Gaussian $\sigma$) and efficiencies for MEG-II}
\newcommand{\m}{\hphantom{$-$}}
\newcommand{\cc}[1]{\multicolumn{1}{c}{#1}}
\begin{tabular}{@{}lll}
\hline
  {\bf PDF parameters }                               & \m  MEG    & \m  MEG-II \\
  \hline\noalign{\smallskip}
  e$^+$  energy  (keV)                                & \m  306 (core)    & \m  130   \\
  e$^+$ $\theta$ (mrad)                               & \m 9.4            & \m  5.3     \\
  e$^+$ $\phi$   (mrad)                               & \m 8.7            & \m  3.7      \\
  e$^+$ vertex (mm) $Z/Y ({\rm core}) $                & \m 2.4 / 1.2      & \m  1.6 / 0.7  \\
  $\gamma$ energy (\%)  ($w<$2\,cm)/($w>$2\,cm)        & \m 2.4 / 1.7      & \m  1.1 / 1.0 \\
  $\gamma$ position (mm) $u/v/w$                       & \m 5 / 5 / 6      & \m  2.6 / 2.2 / 5 \\
  $\gamma$-e$^+$ timing (ps)                          & \m 122            & \m  84 \\ 
  \hline
  {\bf  Efficiency (\%)}                              &                   & \\
  \hline
  trigger                                             & \m $\approx$ 99   & \m $\approx$ 99 \\
  $\gamma$                                            & \m 63             & \m 69 \\
  e$^+$                                               & \m 40             & \m  88  \\ 
\hline
\end{tabular}\\[2pt]
\end{center}
\end{table}

\subsection{MEG-II status}

  The design of a single volume drift chamber with full azimuthal coverage has been finalized in a 2 m long, stereo wires only,  low mass chamber.  Such a longer chamber will also have a higher transparency for positrons  to a system of scintillating tiles than the current MEG detector. This will allow a more precise determination of the positron path length that is crucial to improve the timing of the positron track.
  
   Such detector  has to sustain a hit rate larger than  30 kHz/cm$^2$ on its innermost  layer at the nominal muon stopping rate of 7 10$^7$ muon/s. At the same time it has to provide a  point (R) resolution of the order of 100 $\mu$m. To sustain such rate  1200 almost square 7x7 mm$^2$ cells organized in layers with a 8$^o$ stereo angle are foreseen. A signal \muegamma  positron will therefore  cross  on average 1.7 10$^{-3}$ radiation lengths along its path in the chamber volume. 
   
    The current R\&D program has demonstrated   on  small scale prototypes  that the required resolution can be obtained and that integrating  a charge of 0.5C/cm  on a sense wire would not lead to severe ageing problems. 
    
    The procedure for the wiring of the new MEG-II drift chamber is currently optimized along with all its  mechanical  details. The construction should start at the end of 2014. \\
    
    A new  positron timing detector  is in the construction phase,  with the aim of having 600 scintillator tiles read out by SiPM. Beam tests have demonstrated that a 40 ps resolution can be obtained. The tiles configuration would allow several time measurement along the positron path, a reduction of the pile-up and a better geometrical reconstruction of the positron hits. \\
    
    For the liquid Xenon detector an increase  of the active area of the photo-detectors on the front face is foreseen. Large area SiPM  (10x10 mm$^2$) with an extended sensitivity in the ultra-violet range (where the Xenon scintillating light spectrum  peaks)  will replace the current PMTs. \\

 A new readout electronics to cope with four times more channels in the whole apparatus and to preserve the full waveform recording for each channel is under study. It will be  made of a multi-functional digitization boards with both digitization and trigger capabilities.\\
 
 Ancillary devices (that are not yet part of the MEG-II layout)  are under study. A radiative decay counter able to veto the very low momentum positron from RMD and an active target made of 250 $\mu$m  square scintillating fibers \cite{Papa:2014ica,Papa:2014kxa} are considered.

\section{Conclusions}

 The MEG experiment has set  the world best upper limit on the $\BR$(\muegamma)     to be    5.7 10$^{-13}$ at 90 \% C.L. analyzing a data set of 3.6 10$^{14}$  stopped muons on target. More data are going to be analyzed rather soon to reach a sensitivity of  5.0 10$^{-13}$.

  The MEG-II program is well advanced in the development of new sub-detectors, planning to build the new devices in 2015 and to start data-taking in 2016.
   MEG-II will reach in three years nominal data-taking a  sensitivity of   5.0 10$^{-14}$ with the aim of challenging  a larger and larger  number of  models of NP.

\Acknowledgements
  The author thanks his MEG colleagues for the opportunity to give this talk at this Conference. He acknowledges INFN for its support.

\end{document}